\documentclass[a4paper,onecolumn,groupedaddress,showpacs,showkeys]{revtex4}

\usepackage{graphicx}

\begin{document}

\title{Gate control of the tunneling magnetoresistance 
in double-barrier junctions}

\author{J. Peralta-Ramos}
\email{peralta@cnea.gov.ar}
\affiliation{Departamento de F\'isica, Centro At\'omico
Constituyentes, Comisi\'on Nacional de Energ\'ia At\'omica,
Buenos Aires, Argentina}

\author{A. M. Llois}
\affiliation{Departamento de F\'isica, Centro At\'omico
Constituyentes, Comisi\'on Nacional de Energ\'ia At\'omica,
Buenos Aires, Argentina}
\affiliation{Departamento de F\'isica, Facultad de 
Ciencias Exactas y Naturales, Universidad de Buenos Aires, 
Buenos Aires, Argentina}

\date{\today}

\begin{abstract}
We calculate the conductances and the tunneling magnetoresistance (TMR) 
of double magnetic tunnel
junctions, taking as a model example junctions composed of Fe/ZnSe/Fe/ZnSe/Fe (001).  
The calculations are done as a function of the gate voltage applied to the 
in-between Fe layer 
slab. 
We find that the application of a gate voltage to the in-between Fe slab 
strongly affects the junctions' TMR due to the tuning or untuning of conductance 
resonances mediated by quantum well states. The 
gate voltage allows 
a significant enhancement of the TMR, in a more 
controllable way than by changing the thickness of the 
in-between Fe slab. This effect may be useful in the 
design of future spintronic devices based on the TMR effect, 
requiring large and controllable TMR values. 
\end{abstract}


\maketitle

\newpage
\section{Introduction}
\label{intro}

Double-barrier magnetic tunnel junctions (DBMTJs), in which metallic layers 
are inserted in between the semiconducting region of single-barrier MTJs (SBMTJs), 
are nowadays gaining an increasing interest due to their potential advantages 
over SBMTJs in 
spintronic devices based on the tunneling magnetoresistance effect (TMR)$^1$. 
Although the idea of using double-barrier junctions goes back to the work of Zhang 
{\it et al} a decade ago$^2$, these hybrid systems could be 
epitaxially grown only quite recently.

For example, T. Nozaki {\it et al}$~^3$ have recently measured the TMR of 
fully epitaxial Fe/MgO (001) DBMTJs and 
found larger TMR and V$_{1/2}$ values as compared to 
identically-grown 
single-barrier junctions (V$_{1/2}$ is the bias voltage at which the TMR drops to 
half its value at infinitesimal bias voltage). Z. M. Zeng {\it et al}$~^4$  
have also measured large TMR and V$_{1/2}$ values in Co-Fe-B/Al-oxide double 
junctions. Other 
experimental works along the same lines that confirm these features 
include Refs. 5 and 6.  

Besides their importance for spintronics applications, 
DBMTJs are interesting hybrid systems on their own since they exhibit 
richer transport phenomena than conventional tunnel junctions, that are 
currently under intense investigation. Among these phenomena 
we can mention the  
spin-dependent resonant tunneling due to 
{\it quantum well states} (QWS) inside the in-between metallic 
slab (IBS)$^{7,8}$ and 
the spin-filter 
effect$^{9-11}$. Both effects have been theoretically 
investigated using
realistic electronic structure models only recently$^{9,10,12}$. 
The picture that emerges from these theoretical and experimental 
studies is that double-barrier 
junctions have three major properties unavailable in 
single-barrier ones. First, the TMR values of DBMTJs are, in general,   
significantly larger than those of similarly prepared SBMTJs. 
Second, the dependence of the TMR on bias voltage is considerably smaller 
in DBMTJs than in SBMTJs. And third, the differential conductance 
of DBMTJs shows well defined peaks as a function of bias voltage$^{8,13}$. 
The first two features can be qualitatively understood in terms of 
the spin-filter effect$^9$. The origin of this effect is the 
spin-dependent potential introduced by the in-between magnetic 
layers, which quenches the conductance for the antiparallel 
magnetic configuration of the junction. The third property 
is related to resonant tunneling through quantum well states 
that form in the in-between metallic slab$^{6,9,12,13}$.

From the above, it is clear that 
the investigation of the spin-dependent transport properties of this 
kind of junctions is worthwhile. 
In this work we consider, as a model example, Fe/ZnSe (001) double-barrier 
junctions, and calculate their coherent conductances and TMR. We focus 
on the dependence of the conductances 
and of the TMR on the gate voltage applied to the in-between Fe slab. This is 
an aspect that   
has not been, to our knowledge, studied so far. Starting with the 
pioneering calculations of Zhang {\it et al}$~^2$, the 
main reason for using double-barrier junctions has been to enhance the TMR by 
controlling the thickness of the in-between metallic slab. It was 
theoretically shown by several authors$^7$ that the TMR of DBMTJs  
depends on this thickness, essentially due to the tuning 
or untuning of conductance resonances mediated by quantum well states 
(see, for example, 
the recent experimental work of Niizeki {\it et al}$~^8$). 
Although it is nowadays possible to control the differential conductance by varying the thickness of the 
in-between metallic slab with sub-monolayer precision, in this work we propose to 
control the TMR values of DBMTJs in another way, namely, 
by a gate voltage applied to the 
in-between slab. We analyze this possibility within a simple yet 
realistic model. We note that gate-control has been studied in other systems as well, such 
as carbon nanotubes and quantum dots (see, for instance, Ref. 14 and references therein).   

Our main conclusion is that 
the gate voltage is an important degree of freedom to enhance the TMR value of DBMTJs, 
which may be experimentally rather 
easily available. It allows the 
tuning and untuning of the various conductance resonances 
arising in double-barrier junctions (thus 
enhancing or quenching the TMR), in 
a similar way the thickness of the IBS does, but much more easily from the  
practical standpoint. Furthermore, the very large and controllable 
TMR values obtained in gate-biased 
double-barrier junctions are not attainable using conventional single-barrier 
junctions. 

\section{Systems under study and calculation methods}
Our DBMTJs consist of $m$ layers of BCC Fe (001) inserted in between 
2$n$ layers of zinc-blende ZnSe, so that the Fe midlayers are sandwiched by 
$n$ identical ZnSe layers at each side, the whole multilayer (which we will 
call {\it active region} or AR) being sandwiched by 
two semi-infinite BCC Fe (001) electrodes. We fix $n=$2, which 
represents 1.13 nm of ZnSe at each side of the in-between Fe slab (IBS), 
and consider $m=$2, 3, 4, 6, representing 0.574, 0.861, 
1.148, 1.722 nm of in-between iron.
The junctions are periodic in the 
{\it x-y} plane, being {\it z} the transport direction. 
We note that the junctions are fully epitaxial and that 
interface interdifussion is not taken into account. 
 
In the parallel configuration ($P$), the magnetizations 
of all the magnetic regions (electrodes and in-between Fe layers) 
are parallel 
to each other. In the antiparallel configuration ($AP$), 
the electrodes' magnetization remain parallel to each other but   
the 
Fe midlayer's magnetization is antiparallel to them. 
It is important to note that, 
since 
the coercive fields of the electrodes and of the midlayer are 
different (due to their different thicknesses), 
these magnetic configurations are experimentally 
attainable, as has been shown in recent years.

The electronic structure of the junctions is modeled by a 
Slater-Koster$^{15}$ second nearest 
neighbors {\it spd} tight-binding Hamiltonian fitted to {\it ab initio}
band structure calculations for bulk Fe and bulk ZnSe$^{16}$.
To calculate the mixed hoppings 
between the Fe and the (Zn,Se) atoms in the junctions, 
we use Shiba's rules and Andersen's 
scaling law$^{17}$. The Fe {\it d} bands are spin split by 
$\mu J_{dd}$, where $\mu$=2.2 $\mu_B$ is the experimental magnetic 
moment of bulk Fe and $J_{dd}$=1.16 eV is the exchange integral between 
the Fe {\it d} orbitals ($\mu_B$ is Bohr's magneton). With these values 
for $\mu$ and $J_{dd}$, the bulk Fe {\it d} bands' spin splitting 
is very well reproduced.  
We have also checked that the electrodes' 
and the spacer's band 
structures compare very well to FP-LAPW calculations performed 
with the {\it Wien2k} code$^{18}$. The complex band 
structure of ZnSe, that determines which evanescent states inside 
the semiconducting region are able to couple to Bloch states inside 
the electrodos, is also very well reproduced as 
compared to {\it ab initio} calculations$^{19,20}$. 
The conductances and the TMR of Fe/ZnSe simple junctions, as a function of 
the energy of the incident electrons and of the spacer's thickness, 
are also in very good agreement with the first principles results 
of MacLaren {\it et al}$^{20}$. 
In our DBMTJs, the midlayer Fe {\it d} bands'  
splitting is the same as the one corresponding to bulk Fe. This is a 
rather good approximation since several experimental works$^{21}$  
have shown that the magnetic moment of Fe slabs 
approaches that of bulk Fe even for very thin layers. 
When forming 
the junctions, the ZnSe tight-binding on-site energies are rigidly shifted 
to make the Fe Fermi level fall 1.1 eV 
below its conduction band minimum, as indicated in photoemission experiments 
performed on Fe/ZnSe thin films$^{22}$. 
The gate voltage 
applied to the in-between Fe is simulated by a rigid shift of the on-site energies of 
the Hamiltonian describing the isolated in-between Fe slab. We note that 
Lee {\it et al}$~^{23}$ have been able to fabricate double junctions in which 
the in-between metallic slab could be gate-biased, 
so that the DBMTJs that we study here are, in principle, 
experimentally possible.

The ballistic conductances $\Gamma$ are obtained from Landauer's 
formalism expressed in terms of Green's functions$^{24}$. 
The  
Green's function describing the dynamics of an electron inside the active region 
(ZnSe($n$)/Fe($m$)/ZnSe($n$)) 
is given by 
\begin{equation}
G_S^\sigma=[\hat{1}E_F-H_S^\sigma-\Sigma_L^\sigma-\Sigma_R^\sigma-\Sigma_g^\sigma]^{-1}
\label{green}
\end{equation} 
where $\hat{1}$ stands 
for the unit matrix, $E_F$ is the Fermi energy of the system, 
and $H_S^\sigma$ is the active region's Hamiltonian  
($\sigma$ corresponds to the majority or minority spin
channels). We note that $H_S^\sigma$ depends on the applied gate 
voltage $V_g$, since the on-site energies of the Fe atoms of the IBS are 
rigidly shifted by $V_g$. In Eq. (\ref{green}),  
$\Sigma_{L/R}^\sigma$ are the self-energies 
describing the interaction of the AR with the left ($L$) and right ($R$) 
electrodes, while $\Sigma_g^\sigma$ are the self-energies due to 
the gate electrode contacted to the in-between Fe slab.  
The self-energies due to the iron electrodes are given by 
\begin{equation}
\Sigma_L^\sigma=H_{LS}^{\dagger} g_L^\sigma H_{LS} \qquad \textrm{and} 
\qquad 
\Sigma_R^\sigma=H_{RS}^{\dagger} g_R^\sigma H_{RS} 
\end{equation}
where $H_{LS}$ and $H_{RS}$ are the tight-binding 
couplings of the active region with 
the electrodes, and $g_{L/R}^\sigma$ are the surface Green's functions for each 
electrode. The surface Green's functions are calculated using a semi-analytical 
method$^{25}$ and are exact within our tight-binding approximation. The 
self-energies $\Sigma_g^\sigma$ are taken as complex parameters 
(wave vector and gate-bias {\it independent}), where 
the real and imaginary parts represent the shifting and the 
broadening (finite life-time) of the energy levels of the IBS due to their 
coupling to the gate electrode. In order to simulate a paramagnetic 
gate electrode whose 
band structure near $E_F$ along the (001) direction 
is similar to that of majority (maj.) Fe electrons but not 
to that of minority (min.) Fe ones (for example, gold), we fix the 
imaginary parts of $\Sigma_g^{maj./min.}$ at the constant (energy-independent) 
values of -0.05 eV and -0.01 eV, 
respectively, corresponding to lifetimes 
$\hbar/(-2Im[\Sigma_g^{maj./min.}])$ of 6.58 fs and 32.9 fs. 
With such values we are assuming that the IBS majority electrons can 
leak to the gate electrode more easily than the minority ones, 
because the latter are more confined due to the band structure mismatch 
between the IBS minority bands and the gate electrode's bands. 
The real parts of $\Sigma_g^{maj./min.}$ are calculated as the 
principal part of the Hilbert transforms of 
$-2Im[\Sigma_g^{maj./min.}]$, in order to ensure 
that $\Sigma_g^\sigma$ are causal (retarded)$^{24}$. 
At $E_F$,  the real parts of $\Sigma_g^{maj./min.}$ are equal 
to 0.12 eV and 0.015 eV, respectively.  

The 
transmission probabilities for the transition from the left to the 
right electrodes, $T^\sigma$, are given by$^{24}$
\begin{equation}
T^\sigma(k_{//},E_F)=Tr~ [\Delta_L^\sigma G_S^\sigma \Delta_R^\sigma 
G_S^{\sigma \dagger}]
\end{equation}
where $\Delta_{L/R}^\sigma=i (\Sigma_{L/R}^\sigma-\Sigma_{L/R}^{\sigma \dagger})$ 
are the hybridization functions of the active region with the ($L$,$R$) electrodes. 
$T^\sigma$ gives the probability that an electron coming from the 
left electrode reaches the right electrode. Processes in which an electron 
enters and leaves the gate electrode are not taken into account. That is to say, 
we calculate the conductance due to electrons that tunnel directly from one 
electrode to the other, through the active region whose electronic structure is 
renormalized by the presence of the gate electrode. 
The conductances are then given by 
$\Gamma^\sigma(E_F)=(e^2/h N_{k_{//}}) \sum_{k_{//}} T^\sigma (k_{//},E_F)$,
where $N_{k_{//}}$ is the total number of wave vectors parallel to the interface 
that we consider in our calculations. The tunneling magnetoresistance 
coefficient is defined 
as TMR=100$\times (\Gamma_P-\Gamma_{AP})/\Gamma_{AP}$ (optimistic definition), 
where $\Gamma_P$ 
and $\Gamma_{AP}$ are the conductances in the $P$ and in the 
$AP$ magnetic configurations, respectively. 
By calculating $\Gamma$ using different numbers  
of parallel-to-the-interface wavevectors 
${\bf k_{//}}=k_x {\bf \hat{x}} + k_y {\bf \hat{y}}$ (recall that the junction is 
periodic in the {\it x-y} plane), we find that a mesh of 
5000 ${\bf k_{//}}$ is enough to reach convergence. More details on the method used 
to calculate the conductances can be found in Ref. 10 (see also Ref. 25). 

In this work, we restrict ourselves to zero temperature, 
to infinitesimal {\it bias} voltage (applied to the Fe electrodes) and 
to the coherent regime. 
We assume that the electron's ${\bf k_{//}}$ and spin   
are conserved 
during tunneling, since the junctions are fully epitaxial and 
the Fe midlayer is thin ($<$ 2 nm) and 
ordered$^{24}$.

\section{Results and discussion}
\label{res}

In Figs. 1-4, we show the conductances (upper panels) in the parallel and antiparallel 
configurations of double-barrier junctions with $n$=2 layers (1.13 nm) and 
$m$=2, 3, 4 and 6 layers (0.574, 0.861, 1.148 and 1.722 nm), 
respectively, together with the corresponding TMR 
values (lower panels), as a function of the applied gate voltage. The first thing 
to note from the upper panels of Figs. 1-4 is the appearance of sharp 
conductance peaks at certain values of the gate voltage. 
These peaks occur for both magnetic 
configurations $P$ and $AP$, and are a signature of resonant 
tunneling through polarized quantum well states inside the 
in-between Fe slab. Essentially, the evanescent states 
inside the ZnSe spacers (coupled to Bloch states inside the 
electrodes) can couple to quantum well states 
confined in the in-between Fe slab, thus producing transmission 
resonances. It is clearly seen that it is possible to 
sweep the quantum well states energy spectrum by sweeping the gate voltage. 
That is to say, by changing the gate voltage what we are doing is 
to move the quantum well states in energy. As already mentioned, something 
similar occurs when changing the thickness of the in-between Fe slab$^7$.
The QWS-mediated resonant tunneling phenomenon is displayed 
by all the DBMTJs that we considered. It is particularly clear in those 
DBMTJs with $m$=3 and 6 layers, where sharp $P$ and $AP$ 
conductance peaks, respectively, are observed.

These conductance resonances have an enormous impact on the TMR, as 
it can be seen from the lower panels of Figs. 1-4. The 
TMR values of the DBMTJs that we consider are to be compared to the 
corresponding one for the single-barrier junction with $n$=2 layers (1.13 nm), 
equal to 25 $\%$. It is seen that, even at $V_g=$0, the TMR of DBMTJs 
is larger than that of the corresponding SBMTJ. 
It is also seen that the TMR of DBMTJs can reach, 
under resonant conditions, extremely large values. Considering  
gate voltage values close to zero, the gain in TMR obtained by 
applying a gate voltage is by a factor of 5, 100, 5, 3, for the 
$m$=2, 3, 4, 6 DBMTJs, respectively. The TMR enhancement is particularly 
large for the DBMTJ with $m$=3 layers, in which the TMR goes from 460 $\%$ 
at zero gate (which is 20 times larger than the TMR of the $n=$2 SBMTJ) 
to 47000 $\%$ at -20 mV (which is almost 2000 times larger than the TMR of the SBMTJ). 
A large TMR increase also 
occurs in the double junction with $m$=4 layers, where the TMR goes from 
230 $\%$ to 1000 $\%$ by applying a gate voltage of 10 mV. The 
main conclusion from this analysis is that the TMR of double-barrier 
Fe/ZnSe junctions depends very strongly on the applied gate voltage, and 
that it can reach extremely large values not possible in single-barrier 
junctions. That is to say, the TMR of DBMTJs is, in general, significantly 
larger than that of SBMTJs, and it can be further enhanced by gate-biasing 
the double junctions. 

Another very interesting feature to note is the occurrence of very 
large negative TMR values (in the optimistic definition, the minimum  
TMR value is -100 $\%$). For example, the DBMTJs with $m$=2 layers 
(Fig. 1) 
has a very sharp $AP$ conductance peak at a gate of -60 mV, which 
produces a TMR value very close to -100 $\%$. The same happens for 
the $m$=6 DBMTJ (Fig. 4), at gate voltages equal to -40 mV and 120 mV. 
This switching of the TMR values from positive to negative and 
vice versa, in conjunction with the 
large TMR enhancements that we obtain as compared to the TMR of 
SBMTJs and of DBMTJs at zero gate voltage, may find a use in future 
spintronic devices requiring large and/or variable TMR values. With respect to 
this, we should mention that such large TMR values, both positive and 
negative, are not so far away from what can nowadays be experimentally 
attained. For example, A. Iovan and coworkers$^{6}$ have very recently 
obtained TMR values as high as 10$^4$ $\%$ in Fe/MgO double-barrier junctions, 
the origin of which is resonant tunneling through spin-polarized Fe QWSs. Another 
possible aplication of gate-biased DBMTJs, making use of the very large TMR values 
attainable with these systems, is as magnetic field sensors based 
on the TMR effect$^{26}$. 
One of the key parameters in these devices is the sensitivity of the TMR values  
on the applied magnetic field, and an increase in the TMR results in an 
increase in sensitivity. We have shown in this work that a viable way to  
obtain extremely large values of the TMR is to use gate-biased double-barrier 
magnetic tunnel junctions. It is expected that, if the value of the gate voltage is 
properly chosen so as to produce a large TMR value, the magnetic field sensitivity 
of the DBMTJ will greatly surpass the current sensitivity values attainable in  
single-barrier junctions$^{26}$. 
 
Although our calculations are not self-consistent 
(i.e. we do not take into account charge transfer effects at the 
junctions' interfaces) and are performed at infinitesimal 
bias voltage, we believe that they still capture 
the essential point of this phenomenon. That is, 
the tuning or untuning of conductance 
resonances due to the shifting of quantum well states, and 
their impact on the tunneling magnetoresistance of double-barrier junctions. We 
think that the application 
of a gate voltage to the in-between metallic layers of a double junction is an issue that 
deserves further theoretical and experimental investigation. Control of the 
TMR by a gate voltage is  
better suited to applications than the fine tuning of the in-between metallic 
thickness, and may result in new functionalities for spintronics applications. 
Furthermore, this property is not restricted to Fe/ZnSe (001) DBMTJs, 
since the appearance of spin-polarized quantum well states 
in thin magnetic slabs 
sandwiched by insulating spacers is a rather general phenomenon$^{6-9,12,13}$.  

\section{Summary}
\label{sum}

Taking as a model example junctions composed of 
Fe/ZnSe (001), we have calculated the coherent, zero-bias conductance and the 
tunneling magnetoresistance of 
double-barrier tunnel junctions, as a function of 
the gate voltage applied to the in-between Fe slab and of its thickness. 
The electronic structure of the junctions and their transport 
properties were realistically calculated. We found that 
the tunneling magnetoresistance of double-barrier junctions is 
strongly dependent on the applied gate voltage, essentially due to 
the tuning or untuning of conductance peaks produced by 
resonant tunneling through quantum well states. The tunneling 
magnetoresistance can be tuned to extremely large values 
by sweeping the gate voltage. This 
feature is displayed by all the DBMTJs that we considered. The calculated 
TMR values of the gate-biased double junctions greatly surpass those 
attainable in single-barrier junctions, as well as those of  
double junctions at zero gate voltage. 
The most 
important qualitative conclusion is that the tunneling magnetoresistance can 
be dramatically enhanced by applying small gate voltages, which is 
more easily accessible than controlling the in-between slab's 
thickness. Furthermore, it is possible to obtain large and 
negative TMR values as well. Since the complex band 
structures of ZnSe (001) and of MgO (001) are very similar to each 
other, we believe that these features should also be observed 
in Fe/MgO (001) double-barrier junctions as well. 
These findings may be useful in 
the design of spintronic devices relying on the tunneling 
magnetoresistance effect, and in consequence further 
theoretical and experimental investigation is desirable. For example, it would be 
very interesting to study the influence of the gate voltage on the 
dependence of the tunneling magnetoresistance on bias voltage, which is 
a critical aspect for applications. 

This work was partially funded by UBACyT-X115, PIP-CONICET 6016, 
PICT 05-33304 and PME 06-117. A. M. Llois belongs to CONICET 
(Argentina).

$^1$ J. Fabian, A. Matos-Abiague, C. Ertler, P. Stano, and 
I. Zuti\'c, Acta Physica Slovaca {\bfseries 57}, 565 (2007) 
(arxiv:0711.1461v1 [cond-mat.mtrl-sci] November 12, 2007)

$^2$ X. Zhang, B-Z. Li, G. Sun, and F-C. Pu, 
Phys. Rev. B {\bfseries 56}, 5484 (1997)

$^3$ T. Nozaki, A. Hirohata, N. Tezuka, S. Sugimoto, 
and K. Inomata, Appl. Phys. Lett. 
{\bfseries 86}, 082501 (2005)

$^4$ Z. M. Zeng, H. X. Wei, L. X. Jiang, G. X. Du, W. S. Zhan, 
and X.F. Han, J. Magn. Magn. Mater. {\bfseries 303}, 219 (2006)

$^5$ F. Montaigne, J. Nassar, A. Vaur\`es, F. Nguyen van Dau, 
F. Petroff, A. Schuhl and A. Fert, Appl. Phys. Lett. {\bfseries 73}, 
2829 (1998); 
S. Ohya, P. N. Hai, and M. Tanaka, Appl. Phys. Lett. {\bfseries 87}, 
012105 (2005)

$^6$ A. Iovan, S. Andersson, Y.-G. Naidyiuk, 
A. Vedyayev, B. Dieny and V. Korenivski, arxiv:0705.2375v1 
[cond-mat.mes-hall] (16 May 2007)

$^7$ A. G. Petukhov, A. N. Chantis, and D. O. Demchenko, 
Phys. Rev. Lett. {\bfseries 89}, 107205 (2002); 
N. Ryzhanova, G. Reiss, F. Kanjouri, and A. Vedyayev, 
Phys. Lett. A {\bfseries 329}, 392 (2004); Y.-M. Zhang, and 
S.-J. Xiong, Physica B {\bfseries 362}, 29 (2005); Z. P. Niu, Z. B. 
Feng, J. Yang, and D. Y. Xing, Phys. Rev. B {\bfseries 73}, 
014432 (2006); M. Chshiev, D. Stoeffler, A. Vedyayev, and 
K. Ounadjela, Europhys. Lett. {\bfseries 58}, 257 (2002)

$^8$ T. Niizeki, N. Tezuka, and K. Inomata, Phys. Rev. Lett. 
{\bfseries 100}, 047207 (2008)

$^9$ J. Peralta-Ramos, A. M. Llois, and 
S. Sanvito, {\it Characteristic curves of Fe/MgO (001) single- and 
double-barrier tunnel junctions} (in preparation)

$^{10}$ J. Peralta-Ramos and A. M. Llois, Phys. Rev. B 
{\bfseries 73}, 214422 (2006)

$^{11}$ J. Peralta-Ramos and A. M. Llois, Physica B  
{\bfseries 398}, 393 (2007)

$^{12}$ Y. Wang, Z.-Y. Lu, X.-G. Zhang, and X. F. Han, 
Phys. Rev. Lett. {\bfseries 97}, 087210 (2006)

$^{13}$ T. Nozaki, N. Tezuka, and K. Inomata, 
Phys. Rev. Lett. {\bfseries 96}, 027208 (2006)

$^{14}$ S. Sahoo, T. Kontos, J. Furer, C. Hoffmann, M. Gr$\ddots{a}$ber,
A.Cottet, and C. Sch$\ddots{o}$nenberger, Nature Phys. {\bfseries 1}, 99 (2005); A. Cottet, and M.-S. Choi, 
Phys. Rev. B {\bfseries 74}, 235316 (2006)

$^{15}$ Richard M. Martin, 
{\it Electronic structure: Basic theory and practical methods} 
(Cambridge University Press, Cambridge, 2004); 
J. C. Slater and G. F. Koster, Phys. Rev. {\bfseries 94}, 
1498 (1954) 

$^{16}$ D. A. Papaconstantopoulos, {\it Handbook of 
the band structure of 
elemental solids} (Plenum Press, New York, 1986); 
R. Viswanatha, S. Sapra, B. Satpati, P.V Satyam, 
B.N Dev, 
and D.D Sarma, cond-mat 0505451 v1, 18 May 2005

$^{17}$ O. K. Andersen, Physica B {\bfseries 91}, 317 (1977); 
J. Mathon, Phys. Rev. B {\bfseries 56}, 11810 (1997) 

$^{18}$ P. Blaha, K. Schwarz, G. Madsen, D. Kvasnicka and 
J. Luitz, {\it Wien2k: An augmented plane wave 
+ local orbitals program for calculating crystal properties} 
(Vienna University of Technology, Vienna, 2001)

$^{19}$ P. H. Mavropoulos, N. Papanikolaou, and P. H. Dederichs, 
Phys. Rev. Lett. {\bfseries 5}, 1088 (2000)

$^{20}$ J. M. MacLaren, X.-G. Zhang, W. H. Butler, and X. Wang,
Phys. Rev. B {\bfseries 59}, 5470 (1999); W. H. Butler, X.-G. Zhang, 
T. C. Schulthess, and J. M. MacLaren, Phys. Rev. B 
{\bfseries 63}, 054416 (2001)

$^{21}$ M. Marangolo, F. Gustavsson, M. Eddrief, Ph. Sainctavit, V. H. Etgens,  
V. Cros, F. Petroff, J. M. George, P. Bencok, and N. B. Brookes, Phys. Rev. Lett. 
{\bfseries 88}, 217202 (2002) and references therein.

$^{22}$ M. Eddrief, M. Marangolo, S. Corlevi, G.-M. Guichar, 
V. H. Etgens, R. Mattana, D. H. Mosca and F. Sirotti, 
Appl. Phys. Lett. {\bfseries 81}, 4553 (2002)

$^{23}$ J. H. Lee, K.-I. Jun, K.-H. Shin, S.Y. Park, J.K. Hong, K. Rhie and 
B.C. Lee, J. Magn. Magn. Mater. {\bfseries 286}, 138 (2005)

$^{24}$ Supriyo Datta, {\it Electronic transport in mesoscopic 
systems} (Cambridge University Press, Cambridge, 1999); 
C. Caroli, R. Combescot, P. Nozieres and D. Saint-James, 
J. Phys. C: Solid St. Phys. {\bfseries 4}, 916 (1971); H. Haug and 
A.-P. Jauho, {\it Quantum kinetics in transport and optics of 
semiconductors} (Springer, Germany, 1996)

$^{25}$ S. Sanvito, C. J. Lambert, J. H. Jefferson, and
A. M. Bratkovsky, Phys. Rev. B {\bfseries 59}, 11936 (1999)

$^{26}$ R. C. Chaves, P. P. Freitas, B. Ocker, and W. Maass,
Appl. Phys. Lett. {\bfseries 91}, 102504 (2007) 

\newpage

\begin{figure}[h]
\caption{Conductances ({\it upper panel}) and TMR ({\it lower panel}) 
of a double-barrier junction with $n$=2 (1.13 nm) and $m$=2 (0.574 nm), 
as a function 
of the applied gate voltage.}
\end{figure}

\begin{figure}[h]
\caption{Conductances ({\it upper panel}) and TMR ({\it lower panel}) 
of a double-barrier junction with $n$=2 (1.13 nm) and $m$=3 (0.861 nm), 
as a function 
of the applied gate voltage. The TMR peak at $V_g=$-20 mV has been reduced 
by a factor of 10 in order to fit in the graph.}
\end{figure}

\begin{figure}[htb]
\caption{Conductances ({\it upper panel}) and TMR ({\it lower panel}) 
of a double-barrier junction with $n$=2 (1.13 nm) and $m$=4 (1.148 nm), 
as a function 
of the applied gate voltage.}
\end{figure}

\begin{figure}[htb]
\caption{Conductances ({\it upper panel}) and TMR ({\it lower panel}) 
of a double-barrier junction with $n$=2 (1.13 nm) and $m$=6 (1.722 nm), 
as a function 
of the applied gate voltage.}
\end{figure}

\end{document}